\documentclass{article}

\usepackage[english]{babel}

\usepackage[letterpaper,top=2cm,bottom=2cm,left=3cm,right=3cm,marginparwidth=1.75cm]{geometry}

\usepackage{graphicx}
\usepackage[colorlinks=true, allcolors=blue]{hyperref}
\usepackage{amsfonts,latexsym}
\usepackage{amsmath}
\usepackage{microtype}
\usepackage{booktabs}
\usepackage[dvipsnames]{xcolor}
\usepackage{tabularx}
\usepackage{indentfirst} 
\usepackage{multicol}
\usepackage{pdfpages}

\usepackage{multirow}
\usepackage{listings}
\usepackage{graphicx,mwe}
\usepackage{xcolor}
\usepackage{subcaption}
\usepackage{algorithm,algorithmic}

\begin{document}

\title{State-of-the-Art Approaches to Enhancing Privacy Preservation of Machine Learning Datasets: A Survey}
\date{} 

\author{Chaoyu Zhang, Shaoyu Li}

\maketitle

\begin{abstract}
This paper examines the evolving landscape of machine learning (ML) and its profound impact across various sectors, with a special focus on the emerging field of Privacy-preserving Machine Learning (PPML). As ML applications become increasingly integral to industries like telecommunications, financial technology, and surveillance, they raise significant privacy concerns, necessitating the development of PPML strategies. The paper highlights the unique challenges in safeguarding privacy within ML frameworks, which stem from the diverse capabilities of potential adversaries, including their ability to infer sensitive information from model outputs or training data. We delve into the spectrum of threat models that characterize adversarial intentions, ranging from membership and attribute inference to data reconstruction. The paper emphasizes the importance of maintaining the confidentiality and integrity of training data, outlining current research efforts that focus on refining training data to minimize privacy-sensitive information and enhancing data processing techniques to uphold privacy. Through a comprehensive analysis of privacy leakage risks and countermeasures in both centralized and collaborative learning settings, this paper aims to provide a thorough understanding of effective strategies for protecting ML training data against privacy intrusions. It explores the balance between data privacy and model utility, shedding light on privacy-preserving techniques that leverage cryptographic methods, Differential Privacy, and Trusted Execution Environments. The discussion extends to the application of these techniques in sensitive domains, underscoring the critical role of PPML in ensuring the privacy and security of ML systems.
\end{abstract}
\section{Introduction}
 Recent developments in the field of machine learning, especially in the domain of deep learning, have markedly impacted numerous sectors, such as next-g network\cite{yuaaka, yixin1, yixin2}, financial technology\cite{li2023bijack}, marketing\cite{sung2025advanced,sung2025community}, and surveillance systems\cite{du2022mobile, du2023ucblocker, yu2024pri}, signaling a transformative phase in these industries. Simultaneously, the surge in machine learning-powered artificial intelligence has brought privacy issues to the forefront. Hence, the notion of Privacy-preserving Machine Learning (PPML) has arisen as a prominent area of interest in both the industrial and academic communities.

It is essential to acknowledge that the task of maintaining privacy in machine learning presents distinct challenges, which vary based on the capabilities of potential adversaries. The growing emphasis on understanding and mitigating information leakage from training data involves considering a spectrum of threat models. These models reflect various levels of adversarial power, including the ability to view model outputs, gain access to model parameters, or use certain iterative optimization techniques.   The objectives of these adversaries can vary, such as \emph{membership inference}, \textit{attribute inference}, \textit{property inference}, and \textit{data reconstruction}~\cite{de2021critical,papernot2018sok}. Each of these objectives, as identified in multiple studies, introduces specific difficulties in protecting the integrity and confidentiality of training data within machine learning frameworks.

In the field of PPML, a primary focus is on ensuring that the implemented ML models prevent the escape of confidential information from the training data beyond the trusted boundaries of the data sources. Specifically, during the training phase, the main issues of privacy leakage are centered around the handling of data and its computation. Current research addresses these challenges through two key approaches: (i) determining methods to refine or filter the training data with the aim of either minimizing or entirely eradicating any information sensitive to privacy; and (ii) developing techniques for processing the training data in a way that upholds privacy.

This report concisely examines the attacks associated with \textbf{privacy leakage of training data} in machine learning and delves into the techniques employed for \textbf{preserving training data privacy} in both \textbf{centralized Learning} and \textbf{collaborative learning} to counter these attacks. 
 Through an analysis of these techniques, the report seeks to offer an in-depth understanding of effective strategies for safeguarding training data in machine learning against privacy violations. Additionally, it aims to elucidate the balance between privacy and utility or efficiency that arises in the context of these privacy-preserving approaches.

\noindent \textbf{Privacy-Preserving Techniques.}
Privacy-preserving models in machine learning are fundamentally designed to not disclose extra information about individuals included in their training data. This aspect is of paramount importance for ML systems based on sensitive data, including digital assets\cite{li2023bijack} and medical image processing applications ~\cite{kaissis2021end}. The protection of training data in ML involves a strategic interplay between cryptographic techniques and Differential Privacy (DP). 
Besides, few works utilize a Trusted Execution Environment (TEE) or SGX to achieve PPML ~\cite{hunt2018chiron}.  

\section{A Taxonomy of Adversary Goals}
To compromise data privacy in machine learning systems, adversaries primarily target the exposure of sensitive information embedded in the training dataset. Their goal is to exploit vulnerabilities in machine learning models to extract, reconstruct, or infer private data. This includes conducting Membership Inference Attacks (MIAs) to ascertain if certain data points were used in training, thereby potentially revealing personal or confidential information. Data Reconstruction Attacks take this a step further, attempting to regenerate specific data points or entire datasets from model outputs, thus breaching individual privacy. Property Inference Attacks aim to deduce general properties or characteristics of the training dataset, such as demographic information, without necessarily pinpointing individual data records. Moreover, Model Inversion Attacks specifically attempt to reverse-engineer the model to extract sensitive features of the input data, further endangering data privacy. These attacks underscore the significant risks posed to data privacy by inadequately secured machine learning models, highlighting the necessity for robust privacy-preserving techniques in model training and deployment.

In this paper, We categorize and summarize four different kinds of attacks that are pertinent to the privacy risk associated with the training dataset~\cite{rigaki2023survey}:

\subsection{Membership Inference Attack:} 
The adversary's goal is to determine if a particular record was included in the training dataset of the target 
model~\cite{shokri2017membership}. 

\textbf{Definition: }Given a target machine learning model $ f(x;\theta) $ and a data record $ x $, the goal of MIA is to determine whether $ x $ was part of the training dataset $ D_{train}=\{(x_i, y_i)\}_{i=1}^N $. This can be defined as:

\[
\text{MIA}(f(x;\theta)) = 
\begin{cases} 
1 & \text{if } x \in D_{train}, \\
0 & \text{otherwise}.
\end{cases} 
\]

Here, $ 1 $ indicates that $ x $ is a member of the training set, while $ 0 $ indicates non-membership. MIAs can be categorized based on the attacker's knowledge and access level into black-box and white-box attacks:

\textbf{Black-Box Attack:} In this scenario, the attacker has limited knowledge. They only have access to the model's predictions (output probabilities or classes) for given inputs. The attacker uses this output information to infer whether a particular data point was part of the training dataset. This type of attack is common in real-world scenarios where the internal workings of the model are not exposed\cite{zhang2022label, ye2022enhanced, carlini2022membership}.

\textbf{White-Box Attack:} Here, the attacker has complete knowledge about and access to the target model. This includes the model's architecture, parameters, and training data. White-box attacks are more powerful as they can leverage the inner workings of the model for more accurate inferences about the training data\cite{leino2020stolen}.

Both types of attacks pose significant threats to data privacy, each requiring different defense strategies due to their varied levels of access and knowledge.

\subsection{Data Reconstruction:} Reconstructing samples from the target model's training dataset is the adversary's goal. In this case, a successful attack can result in the training dataset being partially reconstructed~\cite{balle2022reconstructing}.

\textbf{Definition:} Given the training set $D_{\text{train}} = \{(x_i, y_i)\}_{i=1}^N$, where $x_i$ is the image and $y_i \in \{+1, -1\}$ is the label, and a neural network $f(x;\theta)$, we can reconstruct a large subset of the training data by exploiting the implicit biases of neural networks. The neural network aims to solve the max-margin problem: 
$$\underset{\theta'}{\arg \min} \frac{1}{2}\|\theta'\|_{2}^{2} \quad \text{s.t.} \quad \forall i \in [n], y_{i} f_{\theta'}(x_{i}) \geq 1$$
This shows that by optimizing images (and dual parameters) to satisfy the Karush–Kuhn–Tucker (KKT) conditions of the max-margin problem using a trained neural network, training data can be reconstructed. This represents a potential attack leading to the leakage of training data.  A more efficient version of the dataset reconstruction attack, which can provably recover the entire training set in the infinite width regime, is presented in \cite{loo2023understanding,shi2023scale}. In the paper \cite{salem2020updates}, the authors demonstrate that the changes in the output of a black-box ML model, observed before and after an update, can leak information about the dataset used for the update, specifically the updating set.

\subsection{Property Inference:} 
The adversary seeks to uncover sensitive statistical characteristics of the training distribution of the target model~\cite{ganju2018property}.

In the context of property inference attacks, we utilize specific notations and definitions to frame our analysis. Sets are denoted by calligraphic letters, such as $ \mathcal{T} $, while distributions are represented by capital letters, like $ D $. The joint distribution of two random variables, for example, the distribution of labeled instances, is indicated by $ (X, Y) $. Equivalence between two distributions is expressed as $ D_1 \equiv D_2 $. Sampling from a distribution $ X $ is denoted by $ x \leftarrow X $, and the probability of such sampling is indicated by $ \text{Pr}_{x\leftarrow X} $. The support set of a distribution $ X $ is represented by $ \text{Supp}(X) $. We also use the notation $ p \cdot D_1 + (1 - p) \cdot D_2 $ to describe the weighted mixture of two distributions $ D_1 $ and $ D_2 $ with respective weights $ p $ and $ (1 - p) $.

In studying property inference, we adopt the model introduced in prior research. This involves a learning algorithm $ L: (X \times Y)^* \rightarrow H $ that maps datasets in $ \mathcal{T} \in (X \times Y)^* $ to a hypothesis class $ H $. We consider a Boolean property $ f: X \rightarrow \{0, 1\} $ and focus on adversaries who seek to ascertain statistical information about the property $ f $ over the training dataset $ \mathcal{T} $. Specifically, the adversary's objective is to discern whether the fraction of data entries in $ \mathcal{T} $ that exhibit the property $ f $ is $ \hat{f}(\mathcal{T}) = t_0 $ or $ \hat{f}(\mathcal{T}) = t_1 $ for some $ t_0 < t_1 $ within the range [0, 1]. This is achieved in a black-box setting where the adversary can only query the trained model for output labels without access to confidence values or model parameters. This approach aligns with recent explorations in the field of membership-inference attacks.

To formalize property inference attacks, we distinguish between two underlying distributions $ D_-, D_+ $ for dataset instances where $ f(x) = 0 $ and $ f(x) = 1 $, respectively. We then consider two mixed distributions $ D_t \equiv t \cdot D_+ + (1 - t) \cdot D_- $, where $ D_t $ represents the distribution with a $ t $ fraction of points having $ f(x) = 1 $. The adversary's goal is to differentiate between $ D_{t_0} $ and $ D_{t_1} $ for certain values of $ t_0 $ and $ t_1 $, by querying a model $ M $ trained on one of these distributions in a black-box manner. This approach follows methodologies used in previous research and assumes that the adversary can sample from both $ D_- $ and $ D_+ $\cite{mahloujifar2022property, ganju2018property}.

\subsection{Model Inversion and Attribute Inference:}
An adversary having partial knowledge of specific training records and access to a model trained on those records attempts to deduce sensitive information about the records.~\cite{yeom2020overfitting,mehnaz2022your}.

Optimization-based Gradient Inversion and Closed-form Gradient Inversion are two distinct methods used to analyze privacy risks in federated learning, where participants collaboratively train a model by sharing local model gradients or updates.

In Optimization-based Gradient Inversion, an attacker can revert the gradient $ g_i $ uploaded by individual clients $ i \in \{1, 2, \cdots, n\} $ to their respective local datasets $ D_i $ by solving an optimization problem: 
\[ \text{arg min}_{\hat{D}_i} [d(\nabla \hat{D}_i - \nabla D_i) + r(\hat{D}_i)] \]
where $ \hat{D}_i $ refers to randomly initialized dummy samples, $ d() $ is a distance function, and $ r() $ is a regulation function. \cite{zhu2019deep} identified this issue and proposed the deep leakage from gradient (DLG) attack, which uses the second norm as the distance function and the L-BFGS optimizer to solve the optimization problem.

Alternatively, Closed-form Gradient Inversion offers a less computationally demanding method. This technique aims to reconstruct inputs using a direct closed-form derivation, as opposed to an iterative optimization process. The research conducted by Aono et al.~\cite{aono2017privacy}  examines privacy breaches in linear models, revealing that inputs to linear layers can be perfectly reconstructed from their gradients, a situation described as linear leakage. This approach is especially potent in contexts involving linear models, where the relationship between gradients and inputs is directly discernible. This direct method facilitates the recovery of original inputs from gradients in a more straightforward and efficient manner compared to iterative techniques.

Attribute Inference \cite{jia2018attriguard, gong2018attribute} in machine learning and data privacy involves an attacker inferring sensitive attributes ($ y $) of individuals from non-sensitive data ($ x $ and $ z $). Consider a dataset $ D $ with instances as tuples $ (x, y, z) $, where $ x $ and $ z $ are non-sensitive and sensitive attributes, respectively. The attacker leverages a prediction model $ f: X \times Z \rightarrow Y $ trained on $ D $, to construct an inference function $ g: X \times Z \rightarrow Y $ that estimates $ y $. The goal is to maximize the accuracy of $ g $ in predicting $ y $, formulated as $ \text{maximize } \text{Accuracy}(g(x, z), y) $. Various statistical methods, such as logistic regression or neural networks, can be used to build $ g $. This approach highlights privacy risks, especially when $ y $ represents personal data like health or financial information.

\section{Privacy-Preserving Techniques in PPML}
Regarding training data privacy in privacy-preserving machine learning , current approaches \cite{li2024comprehensive} primarily focus on several strategies:
\begin{itemize}
    \item Utilization of \emph{traditional anonymization} mechanisms, such as k-anonymity~\cite{friedman2008providing}, which involves removing identifying information from the training data prior to its use in training processes. This method aims to ensure that individual data points cannot be traced back to specific individuals.
    \item Representation of the original dataset through a surrogate dataset. This is achieved either by grouping anonymized data or abstracting the dataset using \emph{sketch} techniques~\cite{li2019privacy}. These methods aim to retain the utility of the data for analysis and training while reducing the risk of privacy breaches.
    \item Implementation of \emph{differential privacy (DP)} mechanisms~\cite{abadi2016deep,dwork2008differential}. This involves integrating a privacy budget, typically in the form of added noise, into the dataset. The objective is to prevent the leakage of private information by ensuring that the output of the analysis or training does not significantly change when an individual's data is altered or removed, thereby preserving their privacy.
    \item Training ML models on encrypted training data is facilitated by the use of \emph{cryptographic techniques}, notably homomorphic encryption and secure multiparty computation (MPC)~\cite{mohassel2017secureml,liu2021zkcnn}.    
\end{itemize}

From the perspective of computation, current PPML techniques can be categorized into two distinct approaches:
\begin{itemize}
    \item When the training data is processed using traditional anonymization or differential privacy mechanisms, the computational process during training is akin to that in standard, or 'vanilla', model training. In these cases, while the data may be modified or perturbed for privacy reasons, the computational methods and algorithms used for training the model remain largely unchanged from those used in conventional ML training.
    \item In scenarios where the training data is safeguarded through cryptographic methods, the computation involved in privacy-preserving (crypto-based) training becomes more intricate than in regular model training. Cryptographic techniques like homomorphic encryption or secure multiparty computation introduce additional computational complexities. These complexities arise from the need to perform machine learning operations on encrypted data or within the constraints of a secure multi-party computational framework, which often demands more sophisticated and computationally intensive algorithms.
\end{itemize}

\subsection{Differential Privacy}
 In centralized machine learning settings, where a single entity is responsible for training the model, the technique introduced by Abadi et al.~\cite{abadi2016deep}, i.e., DP-SGD, offers enhanced differential privacy protections. DP itself is a key framework that provides a structured way to evaluate the privacy protections of algorithms. 
Differential Privacy (DP) is a technique used to enhance privacy in data processing by adding random noise to training data or model parameters. Informally, it requires that alterations to a single training sample should only have a statistically negligible impact on the output. This use of random noise ensures that the original dataset cannot be reliably deduced from the computation results.

To understand differential privacy more clearly, it's essential to grasp the concept of 'neighboring datasets.' Two datasets, $D$ and $D'$, are considered neighbors if they differ by at most one sample. A randomized algorithm A is said to satisfy $(\epsilon, \delta)-$differential privacy if, for any two neighboring datasets $D$ and $D'$ and for all possible outputs $S$ within the range of $\mathcal{A}$, the following condition is met:
 \begin{equation*}
     \operatorname{Pr}[\mathcal{A}(D) \in S] \leq e^{\epsilon} \cdot \operatorname{Pr}\left[\mathcal{A}\left(D^{\prime}\right) \in S\right]
 \end{equation*}
 
In this formula, $\epsilon$ is a small positive number representing the privacy loss, while $\delta$ is a parameter that gives the probability of this privacy guarantee being broken. The smaller these values, the stronger the privacy guarantee. This inequality ensures that the algorithm's output is similarly likely whether the input dataset is $D$ or $D'$, thus preserving the privacy of individual data points within the dataset.
 Several well-known mechanisms exist for implementing differential privacy, including the exponential mechanism, the Laplacian mechanism, and the Gaussian mechanism~\cite{dwork2014algorithmic}. Each of these mechanisms offers a different way to achieve differential privacy, and the choice between them depends on the specific requirements of the task, such as the type of data and the desired balance between privacy and accuracy.

DP plays a pivotal role in ensuring that the output of an algorithm remains statistically consistent, even if the input data varies by a single record. This concept is particularly crucial in machine learning, where a "record" represents an individual data point in the training set, and the "algorithm" is the machine learning model. DP thus safeguards the privacy of individuals in the dataset by guaranteeing that the inclusion or exclusion of any single data point does not markedly influence the model's overall output. DP-SGD is a specific application of DP in the context of machine learning. It focuses on the manipulation of gradients within the learning algorithm. This involves introducing random perturbations to the gradients calculated during the training process, before they are applied to update the model's parameters. Such perturbations help in maintaining the indistinguishability of individual data points, thereby enhancing privacy.

The general principle of DP-based protocols in PPML involves adding random noise to various elements needing protection, such as training samples, gradients, and model parameters. Applying DP in machine learning ensures that models are resistant to specific types of attacks, such as membership inference attacks (where an attacker tries to determine if a particular data point was used in training) and model inversion attacks (aiming to recreate input data from model outputs).
While differential privacy offers substantial privacy benefits and can be efficient, it often comes with a trade-off in terms of the usability or accuracy of the data. This is due to the noise added for privacy preservation.

In exploring DP-based PPML protocols, it is essential to consider two key aspects: Model Training and Model Prediction.
\subsubsection{DP in Model Training} In this phase, DP can be implemented to ensure that the training process does not reveal sensitive information. This is typically achieved by adding noise to the gradients during the learning process, thus preserving the privacy of individual data points. 

Shokri and Shmatikov's work~\cite{shokri2015privacy} was a significant contribution to the field of privacy-preserving machine learning, particularly in the context of distributed deep networks. They introduced a framework known as Differentially Private Stochastic Gradient Descent (DSSGD) for distributed deep learning. The key feature of this framework was the addition of noise directly to the gradients during the training process. By ensuring that each step of the gradient descent was differentially private, they could guarantee that the final output model also adhered to the principles of differential privacy.

Yu et al.~\cite{yu2019differentially} further advanced this area by presenting a differentially private method specifically tailored for training neural networks. Their approach was notable for its analysis of the privacy loss in differentially private SGD, based on the concept of Concentrated Differential Privacy (CDP)~\cite{dwork2016concentrated}. Concentrated Differential Privacy offers a more refined understanding of privacy loss, allowing for better control over the trade-off between privacy and accuracy in machine learning models.

 Augenstein et al. ~\cite{augenstein2019generative} introduced an innovative approach called Fed-Avg GAN, which applied the federated averaging technique to train GANs in a decentralized setting. Their method was particularly noteworthy for providing user-level differential privacy guarantees. This means that the privacy of individual users' data was protected, even while leveraging the collective data for training more robust and diverse generative models. These advancements demonstrate the evolving landscape of privacy-preserving techniques in machine learning, particularly in complex models like deep networks and GANs.

\subsubsection{DP in Model Prediction} When it comes to model prediction or inference, DP helps ensure that the model's predictions do not compromise individual privacy. This can involve adding noise to the outputs of the model or to the queries made to the model, preventing potential privacy breaches based on the model's responses.

Differential privacy may be used not only to training data, but also to properly safeguard model parameters, thereby preventing any leakage of the model, as mentioned in the study by Carlini et al.~\cite{carlini2018secret}. Differential privacy guarantees that the acquired parameters of a model do not disclose any sensitive details regarding the training data.

Pathak et al.~\cite{pathak2010multiparty} developed a technique particularly designed for combining classifiers in a distributed environment, where each classifier is trained on datasets held by parties that do not trust each other. Their methodology combined these separately trained classifiers while ensuring a guarantee of differentiated privacy. A noteworthy feature of their approach was that the level of noise introduced for privacy purposes decreased as the size of the smallest dataset included increased. Consequently, when the dataset size decreased, a greater amount of noise was necessary to preserve anonymity, potentially compromising the effectiveness of the combined model.

 Expanding on this, Jayaraman et al. introduced an output perturbation technique, as described in their paper cited as ~\cite{jayaraman2018distributed}. This method significantly enhanced the effectiveness of the methodology mentioned in~\cite{pathak2010multiparty}. Their approach consisted of applying Laplace noise directly to the global model after the training process.
 Unlike Pathak et al.’s method, the noise in Jayaraman et al.’s approach was roughly inversely proportional to the size of the entire dataset rather than the smallest dataset. This adjustment meant that the noise level was more manageable and less likely to significantly degrade the utility of the model, especially in scenarios involving larger datasets.

These studies highlight the ongoing evolution and refinement of differential privacy techniques in distributed learning environments, balancing the twin objectives of maintaining robust privacy protections while ensuring the practical utility of the resulting models.

Papernot et al. ~\cite{papernot2016semi} introduced a novel approach known as Private Aggregation of Teacher Ensembles (PATE), which marked a significant advancement in privacy-preserving machine learning. The PATE framework involves dividing sensitive data into multiple disjoint subsets, with each subset used to train a separate "teacher" model. For making predictions, noise is added to the aggregated results of these teacher models. This noise addition is a crucial step in ensuring differential privacy. The labels generated by the teacher models are then used to train a "student" model, which is ultimately deployed for predictions. This approach effectively leverages the strengths of multiple models while maintaining privacy through noise addition and aggregation.

Building on the PATE framework, adaptations have been made for generative models, leading to the development of PATE-GAN~\cite{jordon2018pate} and G-PATE ~\cite{long2019scalable}. These adaptations apply the principles of PATE to generative adversarial networks (GANs), allowing for the generation of synthetic data that retains the statistical properties of real data while preserving privacy.

However, a challenge encountered with G-PATE was the high privacy cost associated with managing high-dimensional gradients. This issue arises because the complexity and dimensionality of the data directly impact the amount of noise required to maintain differential privacy, often leading to a decrease in the utility of the generated data.

To address this, Chen et al.~\cite{chen2020gs} proposed Gradient-Sanitized Wasserstein Generative Adversarial Networks (GS-WGAN). This approach focuses on generating high-dimensional data while still providing a differential privacy guarantee. GS-WGAN applies a method of gradient sanitization to maintain privacy, making it effective for use with both centralized and decentralized datasets. This development signifies a step forward in the ability to generate complex, high-dimensional data in a privacy-preserving manner, demonstrating the evolving capabilities of differential privacy techniques in machine learning.

\subsection{Cryptographic Techniques}
Training ML models on encrypted data is gaining recognition as an effective strategy for safeguarding the privacy of training data. This approach offers a significant advantage over traditional anonymization and DP mechanisms, which are still vulnerable to inference or de-anonymization attacks, particularly when an adversary possesses additional background knowledge. Encryption-based methods provide more robust privacy guarantees by protecting either the training data or the transferred model parameters using cryptosystems. These methods enable computations to be carried out beyond the trusted scope of the data sources, while still maintaining the confidentiality of the data\cite{yu2021fpga, fan2019gpu, shi2023ms}. This growing interest in encryption-based approaches is a response to their potential to offer stronger and more reliable privacy protections in the field of machine learning\cite{xiao2020session, xiao2023bd, xiao2022decentralized,shi2021challenges}. 

 Cryptographic methods, like Homomorphic Encryption (HE), functional encryption schemes, and Secure Multi-Party Computation (MPC)~\cite{cabrero2021sok,patra2020blaze}, allow for secure operations on encrypted data, ensuring the data remains confidential while being processed; DP, conversely, offers robust privacy guarantees by injecting controlled noise into the data, and ensures that individual data points are protected, even when the aggregated data is utilized for ML model training. The synergy between the cryptographic techniques and DP provides a comprehensive approach to safeguarding privacy in ML, balancing the need for data utility with the imperatives of privacy and security.  
 Homomorphic encryption allows for computations to be performed directly on encrypted data, yielding encrypted results that, when decrypted, match the results of operations performed on the plaintext. This means that data can remain encrypted throughout the entire training process, ensuring its confidentiality. Secure multiparty computation, on the other hand, enables multiple parties to jointly compute a function over their inputs while keeping those inputs private.

 SMPC allows for collaborative computation without revealing the underlying data to any of the participating parties, the main primitives of SMPC systems include Garbled circuits(GC), Oblivious Transfer (OT), and Secret Sharing. Both of these cryptographic methods are crucial for enhancing privacy in ML, enabling training on sensitive data without exposing it. 

Given the large-scale datasets typical in machine learning, directly applying generic SMPC protocols for training is often not feasible due to computational inefficiencies. Consequently, researchers are faced with the challenge of developing specialized and efficient protocols for PPML. These protocols must be capable of supporting both the training and prediction phases of machine learning models, even when dealing with large-scale datasets. This task requires a careful balance between maintaining data privacy and ensuring the computational feasibility of processing extensive data volumes.

Additionally, specialized SMPC primitives like Private Set Intersection (PSI) occupy a unique position in the context of machine learning. PSI extends beyond being merely an application of SMPC; it also serves as a fundamental component for other more advanced applications, such as Federated Learning. In this capacity, PSI plays a critical role in enabling collaborative computation and data analysis while preserving the privacy of individual data sets, particularly vital in scenarios where multiple entities need to work together without fully disclosing their respective data. PSI allows parties to jointly compute common samples from their respective datasets without revealing the original datasets. This capability is vital in data preprocessing, as it ensures that privacy is maintained while enabling collaborative computation or learning from multiple data sources. The development of efficient and specialized SMPC protocols continues to be a significant area of research, aiming to balance the privacy needs with the computational challenges in large-scale machine learning applications.

In applications of HE-based machine learning, the typical process involves data owners encrypting their data before transmitting it to a server. When it comes to tasks like making predictions, the required computations are performed directly on the encrypted data, or ciphertexts. This method allows for data processing and analysis while maintaining the confidentiality of the original data, as the server handling the computations never accesses the plaintext form of the inputs. The results of these computations remain encrypted and can only be decrypted by the party that initiated the query. This ensures that the privacy of the data is maintained throughout the process, although it necessitates an additional decryption step by the querying party.

HE also plays a pivotal role in the realm of SMPC. Without the necessity of a trusted third party, it can be employed to produce multiplicative triples in SMPC protocols. This application is evident in protocols such as ABY~\cite{demmler2015aby}. Additionally, HE is instrumental in constructing oblivious transfer protocols, which are fundamental building blocks of SMPC. Oblivious transfer is crucial for ensuring that while a recipient obtains information from a sender, the sender remains unaware of what specific information the recipient has received.

\subsubsection{Model Training}
Demmler et al.~\cite{demmler2015aby} developed ABY, a mixed-protocol framework that stood out for its efficient conversion between three types of secret sharing: Arithmetic sharing, Boolean sharing, and Yao sharing. Each of these sharing types is suited for different kinds of computations (arithmetic, logical, and oblivious transfer-based operations, respectively). ABY's ability to seamlessly convert between these sharing types made it a versatile tool for secure computation.

However, one of the challenges in ABY was that the conversions between different sharing types necessitated a significant number of Oblivious Transfers (OTs), which could diminish the computational efficiency during the online phase of the protocol.

bIn response to this challenge, an improved version named ABY2.0~\cite{patra2021aby2} was proposed. ABY2.0 enhanced the performance of the original ABY framework by shifting the bulk of OTs to the preprocessing phase. This meant that the time-consuming process of executing OTs did not hinder the efficiency of the actual computation phase. Furthermore, ABY2.0 supported multi-input multiplication operations without increasing the online communication overhead, thereby making it more efficient for complex computations.

SecureML~\cite{mohassel2017secureml} used a distinct strategy, emphasizing a configuration that involves two servers that do not collude with each other. The study presented a technique for ensuring the security of neural network training and inference in a scenario where the adversary is semi-honest, meaning they adhere to the protocol but attempt to get extra information. The system developed by SecureML facilitated cooperation among clients while effectively preventing collusion between servers and clients. The objective of this design was to guarantee the integrity of the computation process while safeguarding the data from any internal attacks within the system.

\par Adopting heterogeneous computing architectures can make defending against adversarial attacks in complex systems more viable and cost-effective\cite{al2024sok,liblock,li2023bijack,wu2025blockchain,wu2024strengthening, yixin2}. By integrating various computing resources such as CPUs, GPUs, and FPGAs, security protocols and cryptographic algorithms can be expedited, improving defense strategies' efficiency and practicality\cite{zhang2021high}. This method optimizes the allocation of computational tasks to the most appropriate processors, enhancing both performance and energy efficiency\cite{fan2019gpu, yu2021fpga}. Consequently, heterogeneous computing is crucial for implementing advanced, affordable security measures in safeguarding against adversarial threats\cite{zhang2019optimization, li2020efficient}.

These developments in secure computation frameworks like ABY, ABY2.0, and SecureML highlight the ongoing efforts to balance computational efficiency with robust security measures in the field of privacy-preserving machine learning and secure multiparty computation.

\subsubsection{Model Prediction}
Liu et al.~\cite{liu2017oblivious} introduced MiniONN, a framework designed for privacy-preserving neural network (NN) predictions. MiniONN's approach involves transforming an existing neural network into an oblivious version, enabling secure predictions. This transformation allows the NN to operate on encrypted data, ensuring privacy while maintaining the functionality of the network.

Deepsecure~\cite{rouhani2018deepsecure}, another significant work in this field, also focuses on the prediction phase of oblivious neural networks. It is particularly noted for its scalability and provable security within the semi-honest adversary model, where participants are assumed to follow the protocol but may attempt to glean additional information. Deepsecure, however, relies entirely on Garbled Circuits, a cryptographic technique for secure function evaluation. While GC-based frameworks like Deepsecure are highly secure, they also tend to have high communication costs due to the nature of GC operations.

Addressing the drawbacks of approaches that heavily rely on Garbled Circuits (GC), Riazi et al. in their work "Chameleon"~\cite{riazi2018chameleon} proposed an innovative two-party framework, specifically optimized for efficiency within the context of the semi-honest adversary model. Chameleon distinctively incorporates a semi-honest third party during the offline phase, tasked with generating correlated randomness. This aspect is vital for enhancing the framework's efficiency, offering a novel approach to improving performance in secure computation scenarios.
However, this approach has a potential risk: if the STP colludes with either of the two main parties, some information may be compromised.

When comparing performance and communication efficiency, Chameleon shows significant improvements over both MiniONN and Deepsecure. Specifically, it achieves a 4.2x performance improvement compared to MiniONN, and requires 75x less communication than Deepsecure for running CNNs. These advancements demonstrate the ongoing efforts in the field to optimize privacy-preserving neural network predictions, balancing the trade-offs between security, communication efficiency, and computational performance.

\subsection{Hardware-protected Method}
Trusted Execution Environments (TEEs) like Intel SGX are engineered to enable the secure execution of programs within hardware-protected enclaves. Intel SGX specifically forms secure memory regions within the address space, segregating these areas from all other system codes. This technique facilitates both authentication and encapsulation, allowing external entities to verify that communications originate from an authentic enclave. Within the framework of Privacy-Preserving Machine Learning (PPML) protocols that utilize TEEs, there is an essential process whereby involved parties must transfer confidential data or models into an enclave through a protected channel. Following the authentication of the software operating within the enclave, it can securely process and return encrypted results. These TEE-based methods are relevant and useful in various aspects of machine learning, encompassing both the training phase and the prediction stage of models, as indicated in~\cite{gu2018securing,hunt2018chiron,narra2019privacy}. The use of TEEs in PPML ensures that the computation of sensitive data or models is performed in a secure and isolated environment, thus enhancing privacy and security \cite{wu2024accelerating}.

\textbf{Example.} \cite{hunt2018chiron} introduces Chiron, a system enabling users to train machine learning models on their data while keeping it concealed from the service provider, typified as ML-as-a-Service. Chiron employs hardware-protected SGX enclaves to segregate the service provider's code from the platform, thereby mitigating the risk of data leakage. Additionally, it utilizes a sandbox to restrict the service provider's code, limiting its interaction with the machine learning toolchain and the parameter server.

\begin{figure}[h]
    \centering
    \includegraphics[width=0.5\linewidth]{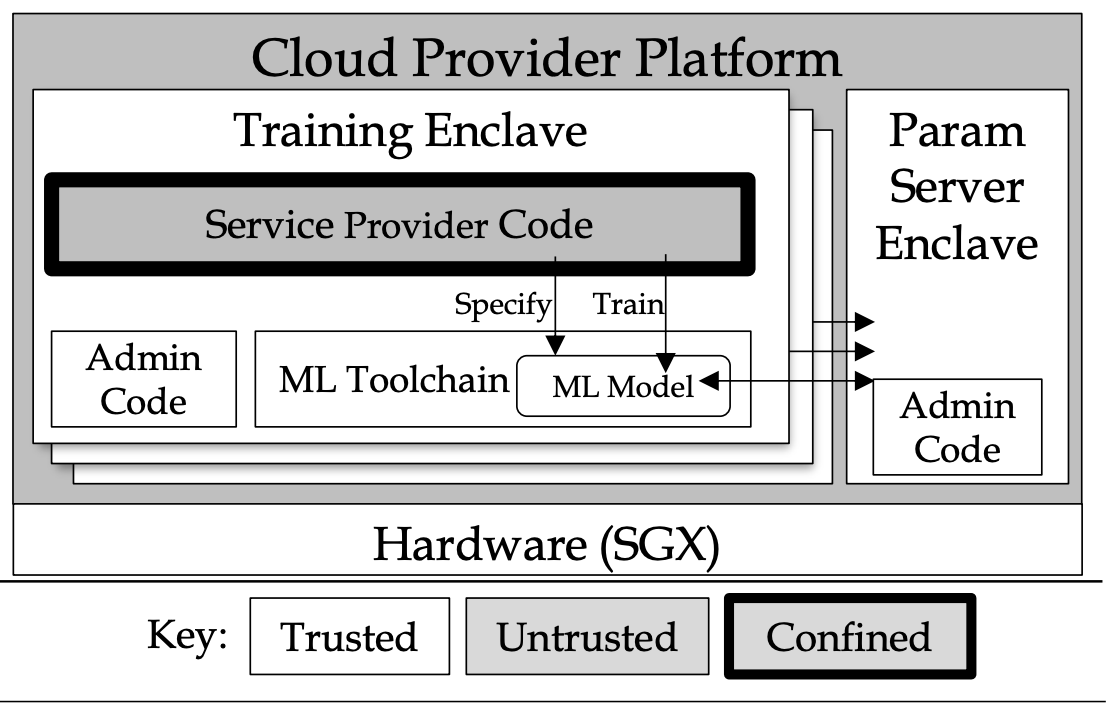}
    \caption{Chiron --- A SGX-based PPML Architecture}
    \label{fig:chiron}
\end{figure}

Figure~\ref{fig:chiron} illustrates the structural design of Chiron. In this system, the service provider's code is confined within multiple training enclaves. Within these enclaves, it can interact via a regulated interface with a trusted ML toolchain. Each enclave contains trusted administrative code, responsible for tasks such as establishing secure communication channels, controlling access to external resources, and loading and confining the service provider's code in a sandbox.
Once the training process is complete, Chiron generates an encrypted model, with the encryption key known exclusively to the user. The service provider has the option to either transmit this model to the user or retain it. If retained, the model is activated within a specific query enclave. In this scenario, the user must utilize Chiron to submit encrypted queries and to receive the outputs from the model.

\section{Defenses in Collaborative/Federated Learning}
While Federated Learning (FL) contributes to privacy, it does not entirely prevent data exposure. Studies have shown that inversion attacks might reveal detailed images from models' weights or gradients~\cite{fredrikson2015model,wang2019beyond,wang2022flare, zhang2023mindfl,hermes}. Additionally, when models are involved in MLaaS, they are vulnerable to misuse or theft by untrustworthy entities. To overcome these shortcomings, FL needs to be enhanced with additional privacy-preserving methods. Implementing FL with secure aggregation (SecAgg) or integrating differential privacy (DP) can impede attacks aimed at dataset reconstruction. Furthermore, applying secure multi-party computation (SMPC) protocols can safeguard the models during the inference process.

\subsection{DP in Federated Learning}
In the realm of Federated Learning (FL), a prominent method for defining functions resistant to adversarial inferences is DP, which effectively limits the potential privacy loss of individual data subjects through the introduction of noise. Within FL's framework, there are two main variants of DP: Local Differential Privacy (LDP) and Central Differential Privacy (CDP). LDP involves each participant adding noise to their updates prior to transmission to the server, whereas CDP involves the server implementing a DP-based aggregation algorithm to ensure data privacy~\cite{naseri2020local}. 
\subsection{Cryptographic Techniques}
Besides, numerous techniques focused on privacy enhancement have been developed to enable various parties to collaboratively train machine learning models while ensuring their private data is not disclosed in its original form. The aforementioned cryptographic techniques have also consistently been integrated into FL to preserve data privacy, e.g., HE and MPC. The HE ensures that the server can decrypt only the combined or aggregated data, i.e., secure aggregation (SecAgg) ~\cite{bonawitz2017practical,kaissis2021end}. 
SMPC enables multiple entities to collaboratively compute a function over their inputs without revealing those inputs to each other.

Private Set Intersection (PSI) is a cryptographic technique that allows two parties, P1 and P2, each holding a private set ($X$ and $Y$, respectively), to determine the intersection of these sets ($X \cap Y$) without revealing any additional information about the individual sets. The only information disclosed through the PSI computation is the intersection itself, ensuring the privacy of the elements outside the intersection.
PSI is particularly relevant in the context of machine learning, especially in vertical federated learning scenarios. Vertical federated learning involves multiple parties collaborating to train a model, where each party holds different features of the same set of samples. PSI is used here to identify the common samples across the parties’ datasets. This process ensures that the model training is conducted only on the overlapping data points, thereby maintaining the privacy of each party's unique data.

Lu and Ding~\cite{lu2020multi} extended the application of PSI to multi-party scenarios within vertical federated learning. Their notable contribution was to accommodate the possibility of participants dropping out during the protocol execution, a practical consideration in real-world applications. This flexibility makes their approach more robust and adaptable to various collaborative environments.

Further advancements in the field have seen the integration of PSI with privacy-preserving biometric searches. For instance, recent developments~\cite{uzun2021fuzzy} have focused on a specialized form of PSI known as fuzzy-labeled PSI, which has been applied to facial search applications. In this context, fuzzy-labeled PSI allows for the matching of biometric data (like facial images) while preserving the privacy of the individuals in the dataset. This approach is particularly useful for sensitive applications where biometric data needs to be matched or searched without compromising individual privacy.

The role of PSI in the field of machine learning, underscores its importance as a tool for safeguarding privacy in collaborative tasks that involve sensitive data processing. This highlights PSI's critical function in balancing the need for data utility with the imperative of maintaining confidentiality, particularly in contexts where multiple parties collaborate on datasets with overlapping but distinct data points.

\subsection{Hybrid Methods: DP + Crypto}
~\cite{kaissis2021end} proposed an FL framework of privacy-preserving medical image processing that incorporates both DP and SecAgg techniques to prevent from model inversion attacks. This design enhances the FL framework's ability to protect sensitive medical data, ensuring that individual records remain confidential while allowing for the collaborative development of robust medical image processing models. PriMIA integrates DP at each FL node, i.e., LDP, to offer assurances of privacy at the patient level. It manages the privacy budget for each node through the use of a Rényi Differential Privacy (RDP) Accountant~\cite{mironov2019r}. This method allows for precise tracking and control of privacy loss incurred during the learning process, ensuring that the privacy guarantees remain within acceptable bounds for each participating node. In the context of training, SMPC is applied to SecAgg the updates of network weights. This is accomplished using additive secret sharing, a technique which is part of the SPDZ protocol~\cite{cryptoeprint:2020/521}, ensuring that each party's contribution to the model update remains confidential while allowing for the collective update to be computed.

\subsection{Measurement and Evaluation of Privacy-Preserving Techniques}
There are many other emerging techniques relevant to preserving training data privacy, e.g., quantification techniques in measuring privacy leakage.

In recent years, the ML privacy field has witnessed a growing trend in the formalization of threat models and adversary capabilities through the use of security and privacy games~\cite{salem2023sok,lukas2023analyzing,wu2016methodology}. This approach, which has been extensively utilized in cryptographic research, involves setting up a security experiment as a probability space. These games set up a security experiment as a probability space, where the effectiveness of an adversary is evaluated based on the probability of their attack succeeding or failing. This approach allows for a more structured and measurable assessment of privacy risks.

In these security and privacy games, various scenarios are constructed to simulate potential attack vectors. The adversary's capabilities are defined within the scope of these scenarios, providing a clear framework for understanding and evaluating the level of threat they pose. By quantifying the likelihood of an adversary's success or failure in these controlled environments, researchers and practitioners can gain a more nuanced and empirical understanding of the vulnerabilities in ML systems.

This methodology not only aids in identifying and measuring privacy risks but also guides the development of more robust and secure ML models. By understanding the probable effectiveness of different types of attacks, ML privacy research can better focus on mitigating those risks that pose the greatest threat to data privacy. This shift towards a more formalized and empirical approach in assessing ML privacy is a crucial step in enhancing the overall security posture of ML systems and protecting sensitive training data from potential breaches.

\section{Comparison and Summary}
To summarize, we provide a comparison among different PPML primitives. 
\begin{enumerate}
    \item SMPC-based (Secure Multi-Party Computation)
    
        \par (a) Performance: Network latency can affect SMPC, and its efficiency is also contingent on the specific protocols or computations being executed.
        \par (b) Privacy: Trust among participants is essential for SMPC to function correctly.
        \par (c) Application: It is typically used in collaborative or distributed learning environments where multiple parties compute a function without revealing their inputs.

    \item ZKP-based (Zero-Knowledge Proofs)

        \par (a) Performance: Zero-Knowledge Proofs aim for smaller proofs to enhance efficiency, and some protocols support parallel computation.
        \par (b) Privacy: The level of privacy varies with the chosen protocol.
        \par (c) Application: ZKPs are often used for data validation processes to prove data integrity without revealing the data itself.

    \item HE-based (Homomorphic Encryption)

        \par (a) Performance: HE can be computationally intensive, and the performance depends on the chosen scheme and parameters.
        \par (b) Privacy: It offers provable security by allowing computations on encrypted data.
        \par (c) Application: HE is ideal for outsourced or centralized computing scenarios, enabling computation on encrypted data stored in the cloud.

    \item FE-based (Functional Encryption)

        \par (a) Performance: The overhead is significantly impacted by the processes of key generation and management.
        \par (b) Privacy: Data access can be fine-grained, but there may be some inevitable leakage of information.
        \par (c) Application: It's used in situations requiring fine-grained access control over encrypted data.

    \item DP-based (Differential Privacy)
    
        \par (a) Performance: The approach is accuracy-limited due to the introduction of noise.
        \par (b) Privacy: Privacy guarantees are noise-dependent, balancing data utility with privacy.
        \par (c) Application: Commonly applied in data analysis and aggregation to protect individual data points within a dataset.
    
    \item TEEs-based (Trusted Execution Environments)

        \par (a) Performance: Overhead is influenced by the setup and management of secure enclaves.
        \par (b) Privacy: Ensures isolation of code and data but may be susceptible to side-channel attacks.
        \par () Application: Used within hardware environments to secure sensitive computations.
\end{enumerate}

Compared to cryptographic mechanisms, DP achieves data security by introducing randomness through the addition of noise, making it an efficient solution for environments with limited computational resources. The implementation of DP in machine learning is relatively less computationally demanding. However, it does impact the model's usability and may decrease prediction accuracy. For practical application, it's necessary to strike a balance between privacy and utility, which can be managed by adjusting the privacy budget.

Beyond measures that secure data during storage and transmission, TEEs provide essential protection for computational processes. However, the reliability of TEEs is significantly contingent on their hardware implementation, which can leave them vulnerable to side-channel attacks.

The selection of each cryptographic primitive is tailored to particular scenarios, guided by the required equilibrium between privacy and utility, the specific use cases, and the computational resources at hand. In the domain of machine learning, these technologies are often integrated, addressing a diverse array of data privacy and security issues. Each primitive involves its own set of trade-offs and is selected based on the unique needs and limitations of the PPML system under development.

\bibliographystyle{abbrv}
\bibliography{sample}

\begin{thebibliography}{10}

\bibitem{abadi2016deep}
M.~Abadi, A.~Chu, I.~Goodfellow, H.~B. McMahan, I.~Mironov, K.~Talwar, and L.~Zhang.
\newblock Deep learning with differential privacy.
\newblock In {\em Proceedings of the 2016 ACM SIGSAC conference on computer and communications security}, pages 308--318, 2016.

\bibitem{al2024sok}
M.~M. Al~Barat, S.~Li, C.~Du, Y.~T. Hou, and W.~Lou.
\newblock Sok: Public blockchain sharding.
\newblock In {\em 2024 IEEE International Conference on Blockchain and Cryptocurrency (ICBC)}, pages 766--783. IEEE, 2024.

\bibitem{aono2017privacy}
Y.~Aono, T.~Hayashi, L.~Wang, S.~Moriai, et~al.
\newblock Privacy-preserving deep learning via additively homomorphic encryption.
\newblock {\em IEEE transactions on information forensics and security}, 13(5):1333--1345, 2017.

\bibitem{augenstein2019generative}
S.~Augenstein, H.~B. McMahan, D.~Ramage, S.~Ramaswamy, P.~Kairouz, M.~Chen, R.~Mathews, et~al.
\newblock Generative models for effective ml on private, decentralized datasets.
\newblock {\em arXiv preprint arXiv:1911.06679}, 2019.

\bibitem{balle2022reconstructing}
B.~Balle, G.~Cherubin, and J.~Hayes.
\newblock Reconstructing training data with informed adversaries.
\newblock In {\em 2022 IEEE Symposium on Security and Privacy (SP)}, pages 1138--1156. IEEE, 2022.

\bibitem{bonawitz2017practical}
K.~Bonawitz, V.~Ivanov, B.~Kreuter, A.~Marcedone, H.~B. McMahan, S.~Patel, D.~Ramage, A.~Segal, and K.~Seth.
\newblock Practical secure aggregation for privacy-preserving machine learning.
\newblock In {\em proceedings of the 2017 ACM SIGSAC Conference on Computer and Communications Security}, pages 1175--1191, 2017.

\bibitem{cabrero2021sok}
J.~Cabrero-Holgueras and S.~Pastrana.
\newblock Sok: Privacy-preserving computation techniques for deep learning.
\newblock {\em Proc. Priv. Enhancing Technol.}, 2021(4):139--162, 2021.

\bibitem{carlini2022membership}
N.~Carlini, S.~Chien, M.~Nasr, S.~Song, A.~Terzis, and F.~Tramer.
\newblock Membership inference attacks from first principles.
\newblock In {\em 2022 IEEE Symposium on Security and Privacy (SP)}, pages 1897--1914. IEEE, 2022.

\bibitem{carlini2018secret}
N.~Carlini, C.~Liu, J.~Kos, {\'U}.~Erlingsson, and D.~Song.
\newblock The secret sharer: Measuring unintended neural network memorization \& extracting secrets.
\newblock {\em arXiv preprint arXiv:1802.08232}, 5, 2018.

\bibitem{chen2020gs}
D.~Chen, T.~Orekondy, and M.~Fritz.
\newblock Gs-wgan: A gradient-sanitized approach for learning differentially private generators.
\newblock {\em Advances in Neural Information Processing Systems}, 33:12673--12684, 2020.

\bibitem{de2021critical}
E.~De~Cristofaro.
\newblock A critical overview of privacy in machine learning.
\newblock {\em IEEE Security \& Privacy}, 19(4):19--27, 2021.

\bibitem{demmler2015aby}
D.~Demmler, T.~Schneider, and M.~Zohner.
\newblock Aby-a framework for efficient mixed-protocol secure two-party computation.
\newblock In {\em NDSS}, 2015.

\bibitem{du2023ucblocker}
C.~Du, H.~Yu, Y.~Xiao, Y.~T. Hou, A.~D. Keromytis, and W.~Lou.
\newblock $\{$UCBlocker$\}$: Unwanted call blocking using anonymous authentication.
\newblock In {\em 32nd USENIX Security Symposium (USENIX Security 23)}, pages 445--462, 2023.

\bibitem{du2022mobile}
C.~Du, H.~Yu, Y.~Xiao, W.~Lou, C.~Wang, R.~Gazda, and Y.~T. Hou.
\newblock Mobile tracking in 5g and beyond networks: Problems, challenges, and new directions.
\newblock In {\em 2022 IEEE 19th International Conference on Mobile Ad Hoc and Smart Systems (MASS)}, pages 426--434. IEEE, 2022.

\bibitem{dwork2008differential}
C.~Dwork.
\newblock Differential privacy: A survey of results.
\newblock In {\em International conference on theory and applications of models of computation}, pages 1--19. Springer, 2008.

\bibitem{dwork2014algorithmic}
C.~Dwork, A.~Roth, et~al.
\newblock The algorithmic foundations of differential privacy.
\newblock {\em Foundations and Trends{\textregistered} in Theoretical Computer Science}, 9(3--4):211--407, 2014.

\bibitem{dwork2016concentrated}
C.~Dwork and G.~N. Rothblum.
\newblock Concentrated differential privacy.
\newblock {\em arXiv preprint arXiv:1603.01887}, 2016.

\bibitem{fan2019gpu}
K.~Fan, C.~Zhang, R.~Shan, H.~Yu, and H.~Jiang.
\newblock Gpu acceleration of ciphertext-policy attribute-based encryption.
\newblock In {\em 2019 20th IEEE/ACIS International Conference on Software Engineering, Artificial Intelligence, Networking and Parallel/Distributed Computing (SNPD)}, pages 94--101. IEEE, 2019.

\bibitem{fredrikson2015model}
M.~Fredrikson, S.~Jha, and T.~Ristenpart.
\newblock Model inversion attacks that exploit confidence information and basic countermeasures.
\newblock In {\em Proceedings of the 22nd ACM SIGSAC conference on computer and communications security}, pages 1322--1333, 2015.

\bibitem{friedman2008providing}
A.~Friedman, R.~Wolff, and A.~Schuster.
\newblock Providing k-anonymity in data mining.
\newblock {\em The VLDB Journal}, 17:789--804, 2008.

\bibitem{ganju2018property}
K.~Ganju, Q.~Wang, W.~Yang, C.~A. Gunter, and N.~Borisov.
\newblock Property inference attacks on fully connected neural networks using permutation invariant representations.
\newblock In {\em Proceedings of the 2018 ACM SIGSAC conference on computer and communications security}, pages 619--633, 2018.

\bibitem{gong2018attribute}
N.~Z. Gong and B.~Liu.
\newblock Attribute inference attacks in online social networks.
\newblock {\em ACM Transactions on Privacy and Security (TOPS)}, 21(1):1--30, 2018.

\bibitem{gu2018securing}
Z.~Gu, H.~Huang, J.~Zhang, D.~Su, A.~Lamba, D.~Pendarakis, and I.~Molloy.
\newblock Securing input data of deep learning inference systems via partitioned enclave execution.
\newblock {\em arXiv preprint arXiv:1807.00969}, pages 1--14, 2018.

\bibitem{hunt2018chiron}
T.~Hunt, C.~Song, R.~Shokri, V.~Shmatikov, and E.~Witchel.
\newblock Chiron: Privacy-preserving machine learning as a service.
\newblock {\em arXiv preprint arXiv:1803.05961}, 2018.

\bibitem{jayaraman2018distributed}
B.~Jayaraman, L.~Wang, D.~Evans, and Q.~Gu.
\newblock Distributed learning without distress: Privacy-preserving empirical risk minimization.
\newblock {\em Advances in Neural Information Processing Systems}, 31, 2018.

\bibitem{jia2018attriguard}
J.~Jia and N.~Z. Gong.
\newblock $\{$AttriGuard$\}$: A practical defense against attribute inference attacks via adversarial machine learning.
\newblock In {\em 27th USENIX Security Symposium (USENIX Security 18)}, pages 513--529, 2018.

\bibitem{jordon2018pate}
J.~Jordon, J.~Yoon, and M.~Van Der~Schaar.
\newblock Pate-gan: Generating synthetic data with differential privacy guarantees.
\newblock In {\em International conference on learning representations}, 2018.

\bibitem{kaissis2021end}
G.~Kaissis, A.~Ziller, J.~Passerat-Palmbach, T.~Ryffel, D.~Usynin, A.~Trask, I.~Lima~Jr, J.~Mancuso, F.~Jungmann, M.-M. Steinborn, et~al.
\newblock End-to-end privacy preserving deep learning on multi-institutional medical imaging.
\newblock {\em Nature Machine Intelligence}, 3(6):473--484, 2021.

\bibitem{cryptoeprint:2020/521}
M.~Keller.
\newblock Mp-spdz: A versatile framework for multi-party computation.
\newblock Cryptology ePrint Archive, Paper 2020/521, 2020.
\newblock \url{https://eprint.iacr.org/2020/521}.

\bibitem{leino2020stolen}
K.~Leino and M.~Fredrikson.
\newblock Stolen memories: Leveraging model memorization for calibrated $\{$White-Box$\}$ membership inference.
\newblock In {\em 29th USENIX security symposium (USENIX Security 20)}, pages 1605--1622, 2020.

\bibitem{li2024comprehensive}
L.~Li.
\newblock Comprehensive survey on adversarial examples in cybersecurity: Impacts, challenges, and mitigation strategies.
\newblock {\em arXiv preprint arXiv:2412.12217}, 2024.

\bibitem{liblock}
L.~Li.
\newblock Mitigating challenges in ethereum's proof-of-stake consensus: Evaluating the impact of eigenlayer and lido.
\newblock {\em arXiv preprint arXiv:2410.23422}, 2024.

\bibitem{li2020efficient}
R.~Li and C.~Zhang.
\newblock Efficient parallel implementations of sparse triangular solves for gpu architectures.
\newblock In {\em Proceedings of the 2020 SIAM Conference on Parallel Processing for Scientific Computing}, pages 106--117. SIAM, 2020.

\bibitem{li2023bijack}
S.~Li, S.~Shi, Y.~Xiao, C.~Zhang, Y.~T. Hou, and W.~Lou.
\newblock Bijack: Breaking bitcoin network with tcp vulnerabilities.
\newblock In {\em European Symposium on Research in Computer Security}, pages 306--326. Springer, 2023.

\bibitem{li2019privacy}
T.~Li, Z.~Liu, V.~Sekar, and V.~Smith.
\newblock Privacy for free: Communication-efficient learning with differential privacy using sketches.
\newblock {\em arXiv preprint arXiv:1911.00972}, 2019.

\bibitem{liu2017oblivious}
J.~Liu, M.~Juuti, Y.~Lu, and N.~Asokan.
\newblock Oblivious neural network predictions via minionn transformations.
\newblock In {\em Proceedings of the 2017 ACM SIGSAC conference on computer and communications security}, pages 619--631, 2017.

\bibitem{liu2021zkcnn}
T.~Liu, X.~Xie, and Y.~Zhang.
\newblock Zkcnn: Zero knowledge proofs for convolutional neural network predictions and accuracy.
\newblock In {\em Proceedings of the 2021 ACM SIGSAC Conference on Computer and Communications Security}, pages 2968--2985, 2021.

\bibitem{long2019scalable}
Y.~Long, S.~Lin, Z.~Yang, C.~A. Gunter, and B.~Li.
\newblock Scalable differentially private generative student model via pate.
\newblock {\em arXiv preprint arXiv:1906.09338}, 2019.

\bibitem{loo2023understanding}
N.~Loo, R.~Hasani, M.~Lechner, A.~Amini, and D.~Rus.
\newblock Understanding reconstruction attacks with the neural tangent kernel and dataset distillation, 2023.

\bibitem{lu2020multi}
L.~Lu and N.~Ding.
\newblock Multi-party private set intersection in vertical federated learning.
\newblock In {\em 2020 IEEE 19th International Conference on Trust, Security and Privacy in Computing and Communications (TrustCom)}, pages 707--714. IEEE, 2020.

\bibitem{lukas2023analyzing}
N.~Lukas, A.~Salem, R.~Sim, S.~Tople, L.~Wutschitz, and S.~Zanella-B{\'e}guelin.
\newblock Analyzing leakage of personally identifiable information in language models.
\newblock {\em arXiv preprint arXiv:2302.00539}, 2023.

\bibitem{mahloujifar2022property}
S.~Mahloujifar, E.~Ghosh, and M.~Chase.
\newblock Property inference from poisoning.
\newblock In {\em 2022 IEEE Symposium on Security and Privacy (SP)}, pages 1120--1137. IEEE, 2022.

\bibitem{mehnaz2022your}
S.~Mehnaz, S.~V. Dibbo, R.~De~Viti, E.~Kabir, B.~B. Brandenburg, S.~Mangard, N.~Li, E.~Bertino, M.~Backes, E.~De~Cristofaro, et~al.
\newblock Are your sensitive attributes private? novel model inversion attribute inference attacks on classification models.
\newblock In {\em 31st USENIX Security Symposium (USENIX Security 22)}, pages 4579--4596, 2022.

\bibitem{mironov2019r}
I.~Mironov, K.~Talwar, and L.~Zhang.
\newblock R$\backslash$'enyi differential privacy of the sampled gaussian mechanism.
\newblock {\em arXiv preprint arXiv:1908.10530}, 2019.

\bibitem{mohassel2017secureml}
P.~Mohassel and Y.~Zhang.
\newblock Secureml: A system for scalable privacy-preserving machine learning.
\newblock In {\em 2017 IEEE symposium on security and privacy (SP)}, pages 19--38. IEEE, 2017.

\bibitem{narra2019privacy}
K.~G. Narra, Z.~Lin, Y.~Wang, K.~Balasubramaniam, and M.~Annavaram.
\newblock Privacy-preserving inference in machine learning services using trusted execution environments.
\newblock {\em arXiv preprint arXiv:1912.03485}, 2019.

\bibitem{naseri2020local}
M.~Naseri, J.~Hayes, and E.~De~Cristofaro.
\newblock Local and central differential privacy for robustness and privacy in federated learning.
\newblock {\em arXiv preprint arXiv:2009.03561}, 2020.

\bibitem{papernot2016semi}
N.~Papernot, M.~Abadi, U.~Erlingsson, I.~Goodfellow, and K.~Talwar.
\newblock Semi-supervised knowledge transfer for deep learning from private training data.
\newblock {\em arXiv preprint arXiv:1610.05755}, 2016.

\bibitem{papernot2018sok}
N.~Papernot, P.~McDaniel, A.~Sinha, and M.~P. Wellman.
\newblock Sok: Security and privacy in machine learning.
\newblock In {\em 2018 IEEE European Symposium on Security and Privacy (EuroS\&P)}, pages 399--414. IEEE, 2018.

\bibitem{pathak2010multiparty}
M.~Pathak, S.~Rane, and B.~Raj.
\newblock Multiparty differential privacy via aggregation of locally trained classifiers.
\newblock {\em Advances in neural information processing systems}, 23, 2010.

\bibitem{patra2021aby2}
A.~Patra, T.~Schneider, A.~Suresh, and H.~Yalame.
\newblock $\{$ABY2. 0$\}$: Improved $\{$Mixed-Protocol$\}$ secure $\{$Two-Party$\}$ computation.
\newblock In {\em 30th USENIX Security Symposium (USENIX Security 21)}, pages 2165--2182, 2021.

\bibitem{patra2020blaze}
A.~Patra and A.~Suresh.
\newblock Blaze: blazing fast privacy-preserving machine learning.
\newblock {\em arXiv preprint arXiv:2005.09042}, 2020.

\bibitem{riazi2018chameleon}
M.~S. Riazi, C.~Weinert, O.~Tkachenko, E.~M. Songhori, T.~Schneider, and F.~Koushanfar.
\newblock Chameleon: A hybrid secure computation framework for machine learning applications.
\newblock In {\em Proceedings of the 2018 on Asia conference on computer and communications security}, pages 707--721, 2018.

\bibitem{rigaki2023survey}
M.~Rigaki and S.~Garcia.
\newblock A survey of privacy attacks in machine learning.
\newblock {\em ACM Computing Surveys}, 56(4):1--34, 2023.

\bibitem{rouhani2018deepsecure}
B.~D. Rouhani, M.~S. Riazi, and F.~Koushanfar.
\newblock Deepsecure: Scalable provably-secure deep learning.
\newblock In {\em Proceedings of the 55th annual design automation conference}, pages 1--6, 2018.

\bibitem{salem2020updates}
A.~Salem, A.~Bhattacharya, M.~Backes, M.~Fritz, and Y.~Zhang.
\newblock $\{$Updates-Leak$\}$: Data set inference and reconstruction attacks in online learning.
\newblock In {\em 29th USENIX security symposium (USENIX Security 20)}, pages 1291--1308, 2020.

\bibitem{salem2023sok}
A.~Salem, G.~Cherubin, D.~Evans, B.~K{\"o}pf, A.~Paverd, A.~Suri, S.~Tople, and S.~Zanella-B{\'e}guelin.
\newblock Sok: Let the privacy games begin! a unified treatment of data inference privacy in machine learning.
\newblock In {\em 2023 IEEE Symposium on Security and Privacy (SP)}, pages 327--345. IEEE, 2023.

\bibitem{shi2023scale}
S.~Shi, N.~Wang, Y.~Xiao, C.~Zhang, Y.~Shi, Y.~T. Hou, and W.~Lou.
\newblock Scale-mia: A scalable model inversion attack against secure federated learning via latent space reconstruction.
\newblock {\em arXiv preprint arXiv:2311.05808}, 2023.

\bibitem{shi2023ms}
S.~Shi, Y.~Xiao, C.~Du, M.~H. Shahriar, A.~Li, N.~Zhang, Y.~T. Hou, and W.~Lou.
\newblock Ms-ptp: Protecting network timing from byzantine attacks.
\newblock In {\em Proceedings of the 16th ACM Conference on Security and Privacy in Wireless and Mobile Networks}, pages 61--71, 2023.

\bibitem{shi2021challenges}
S.~Shi, Y.~Xiao, W.~Lou, C.~Wang, X.~Li, Y.~T. Hou, and J.~H. Reed.
\newblock Challenges and new directions in securing spectrum access systems.
\newblock {\em IEEE Internet of Things Journal}, 8(8):6498--6518, 2021.

\bibitem{shokri2015privacy}
R.~Shokri and V.~Shmatikov.
\newblock Privacy-preserving deep learning.
\newblock In {\em Proceedings of the 22nd ACM SIGSAC conference on computer and communications security}, pages 1310--1321, 2015.

\bibitem{shokri2017membership}
R.~Shokri, M.~Stronati, C.~Song, and V.~Shmatikov.
\newblock Membership inference attacks against machine learning models.
\newblock In {\em 2017 IEEE symposium on security and privacy (SP)}, pages 3--18. IEEE, 2017.

\bibitem{sung2025advanced}
P.-H. Sung.
\newblock Advanced machine learning for housing market analysis: Predicting property prices in washington, dc, 2025.

\bibitem{sung2025community}
P.-H. Sung.
\newblock Community structure and connectivity analysis in social networks using community detection methods.
\newblock {\em Authorea Preprints}, 2025.

\bibitem{uzun2021fuzzy}
E.~Uzun, S.~P. Chung, V.~Kolesnikov, A.~Boldyreva, and W.~Lee.
\newblock Fuzzy labeled private set intersection with applications to private $\{$Real-Time$\}$ biometric search.
\newblock In {\em 30th USENIX Security Symposium (USENIX Security 21)}, pages 911--928, 2021.

\bibitem{wang2022flare}
N.~Wang, Y.~Xiao, Y.~Chen, Y.~Hu, W.~Lou, and Y.~T. Hou.
\newblock Flare: defending federated learning against model poisoning attacks via latent space representations.
\newblock In {\em Proceedings of the 2022 ACM on Asia Conference on Computer and Communications Security}, pages 946--958, 2022.

\bibitem{wang2019beyond}
Z.~Wang, M.~Song, Z.~Zhang, Y.~Song, Q.~Wang, and H.~Qi.
\newblock Beyond inferring class representatives: User-level privacy leakage from federated learning.
\newblock In {\em IEEE INFOCOM 2019-IEEE conference on computer communications}, pages 2512--2520. IEEE, 2019.

\bibitem{wu2024accelerating}
K.~W. Wu.
\newblock Accelerating sparse graph neural networks with tensor core optimization.
\newblock {\em arXiv preprint arXiv:2412.12218}, 2024.

\bibitem{wu2024strengthening}
K.~W. Wu.
\newblock Strengthening defi security: A static analysis approach to flash loan vulnerabilities.
\newblock {\em arXiv preprint arXiv:2411.01230}, 2024.

\bibitem{wu2025blockchain}
K.~W. Wu.
\newblock Blockchain-based secure vehicle auction system with smart contracts.
\newblock {\em arXiv preprint arXiv:2501.04841}, 2025.

\bibitem{wu2016methodology}
X.~Wu, M.~Fredrikson, S.~Jha, and J.~F. Naughton.
\newblock A methodology for formalizing model-inversion attacks.
\newblock In {\em 2016 IEEE 29th Computer Security Foundations Symposium (CSF)}, pages 355--370. IEEE, 2016.

\bibitem{xiao2022decentralized}
Y.~Xiao, S.~Shi, W.~Lou, C.~Wang, X.~Li, N.~Zhang, Y.~T. Hou, and J.~H. Reed.
\newblock Decentralized spectrum access system: Vision, challenges, and a blockchain solution.
\newblock {\em IEEE Wireless Communications}, 29(1):220--228, 2022.

\bibitem{xiao2023bd}
Y.~Xiao, S.~Shi, W.~Lou, C.~Wang, X.~Li, N.~Zhang, Y.~T. Hou, and J.~H. Reed.
\newblock Bd-sas: Enabling dynamic spectrum sharing in low-trust environment.
\newblock {\em IEEE Transactions on Cognitive Communications and Networking}, 2023.

\bibitem{xiao2020session}
Y.~Xiao, S.~Shi, N.~Zhang, W.~Lou, and Y.~T. Hou.
\newblock Session key distribution made practical for can and can-fd message authentication.
\newblock In {\em Annual Computer Security Applications Conference}, pages 681--693, 2020.

\bibitem{ye2022enhanced}
J.~Ye, A.~Maddi, S.~K. Murakonda, V.~Bindschaedler, and R.~Shokri.
\newblock Enhanced membership inference attacks against machine learning models.
\newblock In {\em Proceedings of the 2022 ACM SIGSAC Conference on Computer and Communications Security}, pages 3093--3106, 2022.

\bibitem{yeom2020overfitting}
S.~Yeom, I.~Giacomelli, A.~Menaged, M.~Fredrikson, and S.~Jha.
\newblock Overfitting, robustness, and malicious algorithms: A study of potential causes of privacy risk in machine learning.
\newblock {\em Journal of Computer Security}, 28(1):35--70, 2020.

\bibitem{yuaaka}
H.~Yu, C.~Du, Y.~Xiao, A.~Keromytis, C.~Wang, R.~Gazda, Y.~T. Hou, and W.~Lou.
\newblock Aaka: An anti-tracking cellular authentication scheme leveraging anonymous credentials.
\newblock In {\em Network and Distributed System Security Symposium (NDSS)}, 2023.

\bibitem{yu2024pri}
H.~Yu, S.~Shi, Y.~Shi, E.~Burger, Y.~T. Hou, and W.~Lou.
\newblock Pri-share: Enabling inter-sas privacy protection via secure multi-party spectrum allocation.
\newblock In {\em 2024 IEEE International Symposium on Dynamic Spectrum Access Networks (DySPAN)}, pages 347--356. IEEE, 2024.

\bibitem{yu2021fpga}
H.~Yu, C.~Zhang, and H.~Jiang.
\newblock A fpga-based heterogeneous implementation of ntruencrypt.
\newblock In {\em Advances in Parallel \& Distributed Processing, and Applications: Proceedings from PDPTA'20, CSC'20, MSV'20, and GCC'20}, pages 461--475. Springer, 2021.

\bibitem{yu2019differentially}
L.~Yu, L.~Liu, C.~Pu, M.~E. Gursoy, and S.~Truex.
\newblock Differentially private model publishing for deep learning.
\newblock In {\em 2019 IEEE symposium on security and privacy (SP)}, pages 332--349. IEEE, 2019.

\bibitem{hermes}
C.~Zhang, S.~Shi, N.~Wang, X.~Xu, S.~Li, L.~Zheng, R.~Marchany, M.~Gardner, Y.~T. Hou, and W.~Lou.
\newblock Hermes: Boosting the performance of machine-learning-based intrusion detection system through geometric feature learning.
\newblock In {\em Proceedings of the Twenty-fifth International Symposium on Theory, Algorithmic Foundations, and Protocol Design for Mobile Networks and Mobile Computing}, pages 251--260, 2024.

\bibitem{zhang2023mindfl}
C.~Zhang, N.~Wang, S.~Shi, C.~Du, W.~Lou, and Y.~T. Hou.
\newblock Mindfl: Mitigating the impact of imbalanced and noisy-labeled data in federated learning with quality and fairness-aware client selection.
\newblock In {\em MILCOM 2023-2023 IEEE Military Communications Conference (MILCOM)}, pages 331--338. IEEE, 2023.

\bibitem{zhang2021high}
C.~Zhang, H.~Yu, Y.~Zhou, and H.~Jiang.
\newblock High-performance and energy-efficient fpga-gpu-cpu heterogeneous system implementation.
\newblock In {\em Advances in Parallel \& Distributed Processing, and Applications: Proceedings from PDPTA'20, CSC'20, MSV'20, and GCC'20}, pages 477--492. Springer, 2021.

\bibitem{zhang2022label}
G.~Zhang, B.~Liu, T.~Zhu, M.~Ding, and W.~Zhou.
\newblock Label-only membership inference attacks and defenses in semantic segmentation models.
\newblock {\em IEEE Transactions on Dependable and Secure Computing}, 20(2):1435--1449, 2022.

\bibitem{yixin1}
Y.~Zhang, W.~Cheng, J.~Wang, and W.~Zhang.
\newblock Performance analysis and blocklength minimization of uplink rsma for short packet transmissions in urllc.
\newblock In {\em GLOBECOM 2023 - 2023 IEEE Global Communications Conference}, pages 6765--6770, 2023.

\bibitem{yixin2}
Y.~Zhang, W.~Cheng, and W.~Zhang.
\newblock Multiple access integrated adaptive finite blocklength for ultra-low delay in 6g wireless networks.
\newblock volume~23, pages 1670--1683, 2024.

\bibitem{zhu2019deep}
L.~Zhu, Z.~Liu, and S.~Han.
\newblock Deep leakage from gradients, 2019.

\end{thebibliography}

\end{document}